\documentclass[aps,prl,superscriptaddress,showpacs,preprint]{revtex4-1}

\usepackage{amsmath}
\usepackage{amssymb}
\usepackage{amsfonts} 
\usepackage{latexsym}
\usepackage{bbm}
\usepackage{indentfirst} 
\usepackage{graphicx}
\usepackage{subfigure}
\usepackage{mathrsfs}
\usepackage{varioref}
\usepackage{soul}
\usepackage{xcolor}

\usepackage{dcolumn}
\usepackage{bm}
\usepackage{marginnote}
\usepackage{url}

\begin{document}

\preprint{xxxx}

\title{Interplay between the Weibel instability and the Biermann battery in realistic laser-solid interactions}

\author{N. Shukla}
\email{nshukla@ist.utl.pt}
\affiliation{GoLP/Instituto de Plasmas e Fus\~ao Nuclear, Instituto Superior T\'ecnico, Universidade de Lisboa, Lisbon, Portugal}%
\author{K. Schoeffler}%
\email{kevin.schoeffler@tecnico.ulisboa.pt}
\affiliation{GoLP/Instituto de Plasmas e Fus\~ao Nuclear, Instituto Superior T\'ecnico, Universidade de Lisboa, Lisbon, Portugal}%
\author{E. Boella}%
\affiliation{Physics Department, Lancaster University, Lancaster, UK}
\affiliation{Cockcroft Institute, Sci-Tech Daresbury, Warrington, UK}
\author{J.Vieira}%
\affiliation{GoLP/Instituto de Plasmas e Fus\~ao Nuclear, Instituto Superior T\'ecnico, Universidade de Lisboa, Lisbon, Portugal}%
\author{R. Fonseca}%
\affiliation{GoLP/Instituto de Plasmas e Fus\~ao Nuclear, Instituto Superior T\'ecnico, Universidade de Lisboa, Lisbon, Portugal}%
\affiliation{DCTI/ISCTE, Instituto Universitario de Lisboa, Lisbon, Portugal}%
\author{L. O. Silva}%
\email{luis.silva@tecnico.ulisboa.pt}
\affiliation{GoLP/Instituto de Plasmas e Fus\~ao Nuclear, Instituto Superior T\'ecnico, Universidade de Lisboa, Lisbon, Portugal}%

\date{\today}

\begin{abstract}
A novel setup allows the Weibel instability and its interplay with the Biermann battery to be probed in laser-driven collisionless plasmas. \textit{Ab initio} particle-in-cell (PIC) simulations of the interaction of short ($ \le ps$) intense $(a_0 \ge 1)$ laser-pulses with overdense plasma targets show observable Weibel generated magnetic fields. This field strength surpasses that of the Biermann battery, usually dominant in experiments, as long as the gradient scale length is much larger than the local electron inertial length; this is achievable by carefully setting the appropriate gradients in the front of the target e.g. by tuning the delay between the main laser pulse and the pre-pulse.

\end{abstract}

\maketitle

The origin and evolution of magnetic fields starting from initially unmagnetized plasmas is a long-standing question, which has implications not only in astrophysics (e.g. Gamma-ray-bursts, TeV-Blazar, etc) \cite{Uzdensky-2014,Widrow, Kronberg, Kulsrud-2008} but also in laboratory plasmas (e.g. fast ignition) \cite{Luis-2002,Tzoufras-2006, Shukla-JPP-2012}. Magnetic field growth in astrophysical conditions is often attributed to the turbulent dynamo mechanism, which requires an initial seed field. The dominant processes responsible for magnetogenesis, i.e. the generation of these initial fields, are still under strong debate. Among the known mechanisms, the Biermann battery and the Weibel or current filamentation instability are two major candidates \cite{Medvedev-ApJ-1999, Ricardo-POP-1999, Medvedev-ApJ-2005, Nishikawa-2009, Shukla2010,   Bret-APJ-2009, Ruyer-PRL-2017}. The Biermann battery acts in the presence of temperature and density gradients perpendicular to each other \cite{Gruzinov-2001, Brandenburg2012}. In contrast, the Weibel instability is driven by temperature anisotropies \cite{Weibel-1959, Bret-POP-2010}. These key mechanisms have been reproduced using scaled experiments governed by similar physical laws \cite{Tata-2002, Mondal}. The interplay between the Biermann battery effect and the Weibel instability in the laboratory is both of fundamental interest and relevant to understand magnetogenesis.

Recent developments in laser technology (intensities in excess of $10^{19} \mathrm{W/cm^2}$ with laser pulse durations shorter than 1 ps and high-resolution diagnostics) open the possibility to probe such processes through laser-solid interactions \cite{Tata-2002, Ruyer-2015, Fiuza-2012, Boella_2018, PhysRevE.80.027401}. In these experiments, the magnetic field generation is often attributed to the Biermann battery \cite{Stamper-1971, Sakagami1979, Mondal}. The Biermann field grows linearly as $B(t)$ $\approx -(tc/n_e e) \nabla n_e \times \nabla T_e$ $\approx (tc/e)(k_B T_e/L_T L_n)$, where $L_n \equiv n_e/\nabla n_e$ and $L_T \equiv T_e/\nabla T_e$ are the density and temperature gradient scale lengths, respectively, $k_B$ is the Boltzmann constant, $n_e$ and $T_e$ are the electron density and temperature, $e$ is the elementary charge, and $c$ is the speed of light in vacuum. Theoretical and computational studies have demonstrated magnetic field generation via the Biermann battery \cite{cadjan_ivanov_ivlev_1997, Wilks1997} in the context of hydrodynamical systems. Recently, Schoeffler\,\textit{et.al}\, \cite{Schoeffler-2014,Schoeffler2016} investigated the kinetic effects of the Biermann battery in a collisionless expanding plasma, finding that for sufficiently large gradient scale length $L \sim L_n \sim L_T$ the Weibel instability competes with the Biermann battery. The relative importance of the Biermann battery can be adjusted by changing the scale length of the density and temperature gradients. The saturated Biermann battery generated field obeys the scaling:

\begin{equation}
	\frac{B}{\sqrt{8 \pi P_{plasma}}} = \beta_e^{-1/2} \sim \frac{d_e}{L},
\end{equation}
where $P_{plasma}$ is the plasma pressure, $d_e \equiv c/\omega_{p}$ and $\omega_p = (4 \pi e^2 n_e /m_e)^{1/2}$ are the respective electron skin depth and plasma frequency, and $m_e$ is the electron rest mass. For systems where $L/d_e < 100$ the dominant magnetic field is generated via the Biermann mechanism. In contrast when $L/d_e \geq 100$, the Weibel instability generates magnetic fields that are stronger and grow faster than that of the Biermann battery.

In this Letter, we carry out a numerical and theoretical study using particle-in-cell (PIC) simulations to investigate magnetic fields generated by the Weibel instability in the interaction of a short (ps) high intensity ($a_0 \geq 1$) laser pulse and a plasma with sufficiently large $L$. Until now, the large simulation domains and long simulation times required to capture these mechanisms have impeded detailed exploration of this regime. Our simulation results reveal that by tuning the delay between an ionizing pre-pulse and the main pulse, and defining the spot size of the laser such that $L/d_e \geq 100$, the Weibel generated magnetic field magnitude surpasses the usually observed Biermann field, and can be directly observed in current laser-plasma interaction experiments. 

We simulate the interaction of an ultraintense laser pulse with a fully ionized unmagnetized electron-proton plasma with realistic mass ratio (proton mass $m_i = 1836 \, m_e$) using the OSIRIS framework \cite{Fonseca-2002, Fonseca2008, Fonseca2013}. The laser is s-polarized (i.e. the electric field is perpendicular to the simulation plane) and has a peak intensity $I_L = 10^{19}\,\mathrm{W/cm^2}$ (normalized vector potential $a_0 = 2$) with a wavelength $\lambda_{0}= 1.0\,\mu m$. We choose s-polarization to isolate the out-of-plane Biermann and Weibel magnetic fields from the laser field. Furthermore, s-polarization in 2D better approximates 3D conditions, as both conditions have been shown to produce less heating than with p-polarization in 2D \cite{Wilks1997,Chopineau}. We have performed 2D simulations with similar laser parameters using p-polarized laser confirming the conclusion predicted by Ref. \cite{Wilks1997}. We define $\omega_p$ and $d_e$ using a reference plasma density $n_0=1.1\times 10^{22}\, \text{cm}^{-3}=10 \, n_c$, where $n_c = \omega_{0}^2 m_e / 4 \pi e^2$ is the critical density, and $\omega_{0} = 2\pi c/\lambda_{0}$ is the laser frequency. The envelope of the pulse follows a flat-top function having rise (R) and fall (F) time $\tau_R = \tau_F = 10.0 \,\omega_p^{-1}(1.7 \, \text{fs})$ and duration $\tau_{FT} = 1034 \,\omega_p^{-1}(175 \, \text{fs})$. Its transverse profile is modelled as a Gaussian function with spot size at full width half maximum (FWHM) $\text{w}_{FWHM} =100\,d_e (5 \, \mu\text{m})$. These are typical laser parameters in laser-solid interaction experiments \cite{Wei2004}.

The laser (propagating along the $x_1$ direction) interacts with a plasma having longitudinal electron density profile $n_e(x_1) = 0.5 \, n_0 \left \{ \tanh \left[2\left(x_1 - x_{10} \right)/L_n \right] + 1 \right \}$, where $n_{0}=10 \, n_c$ is the maximum density, $x_1$ is the longitudinal coordinate, and $L_n(= n_0/\nabla n_e(x_{10}))$ is the initial density scale length where in our primary simulation $L_n =400 \,d_e (20 \, \mu\text{m})$. The laser focal point coincides with the location of critical density at $x_{10} = 1250 \, d_e$. The electrons and ions have initial temperatures $T_{e0} = 1 \, \text{keV}$ and $T_{i0} = 1 \, \text{eV}$, respectively (small compared to the laser heating, but large enough to resolve the Debye length).

The simulation box size $L_{x1}\times L_{x2} = 2000 \times 2000\,d_e^2$ is divided into $20000 \times 20000$ cells and a time step $\Delta t = 0.05 \, \omega_p^{-1}$.  Each cell contains 12 macro-particles per species, whose dynamics have been followed for more than 100000 time steps. We choose absorbing boundary conditions along $x_1$ and periodic along $x_2$ for fields and particles. Increased transverse box sizes $L_{x2}$, spatial and temporal resolution, and number of particles per cell were tested, showing overall convergence.


\begin{figure}[htp]
\begin{center}
\includegraphics[scale=0.25]{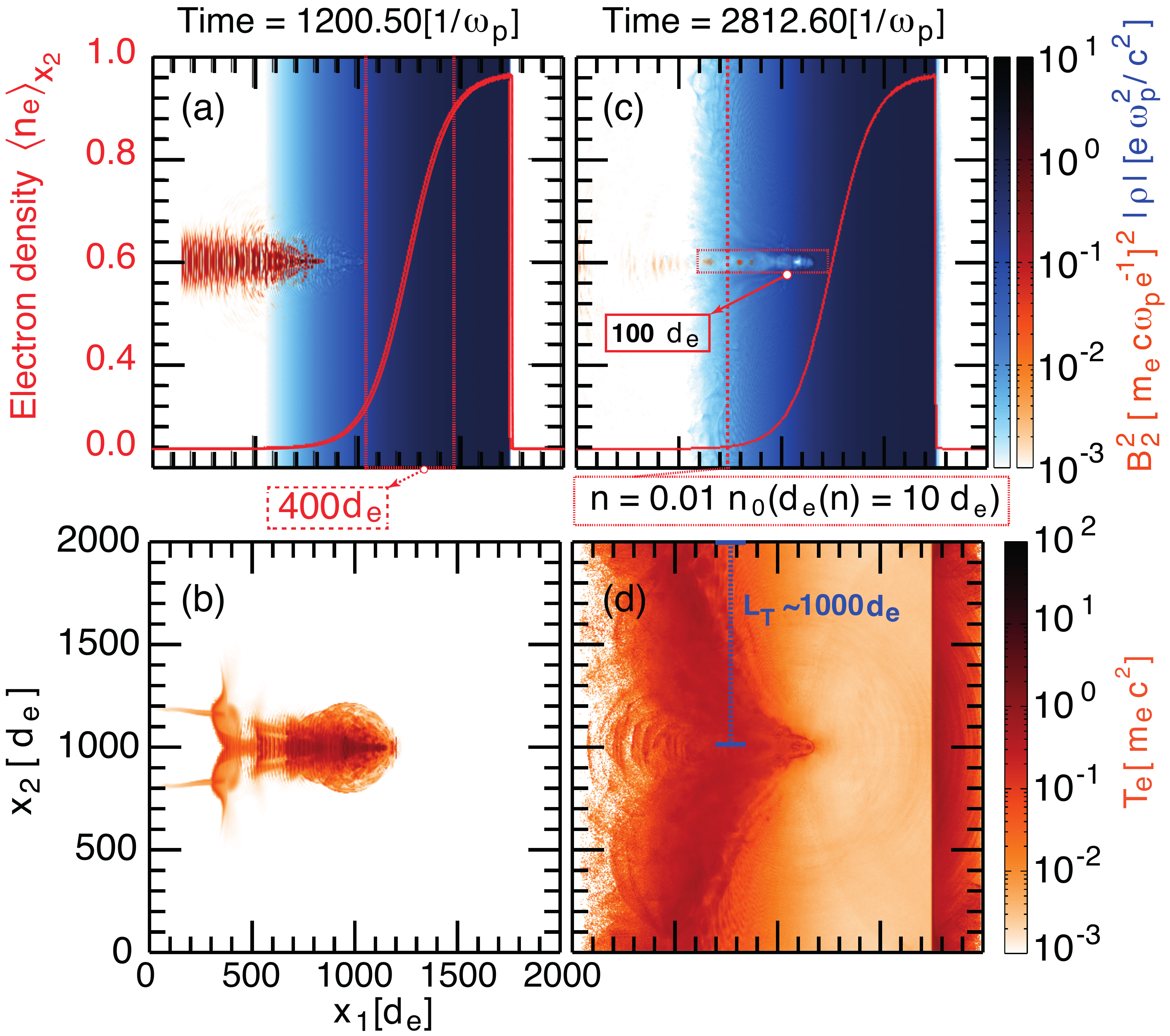}
\caption{Electron density $n_e$ (blue) and laser magnetic field $B_2^2$ (orange) (a, c) and electron temperature $T_e$ (b, d) at $t = 1200.5$ and $2812.6 \, \omega_{p}^{-1}$,  respectively. The red solid lines in (a) and (c) are an average of the density along the $x_2$ direction, and the dashed red line shows the gradient length scale $L_n = 400 \, d_e$. The red dashed box in (c) indicates the focal spot of the laser. The red dashed line defines the boundary between $ L_n > 100 \, d_e(n_e)$ (left), and $ L_n < 100 \, d_e(n_e)$ (right). The blue dashed lines in (d) point to the location where $L_T \simeq 1000 \, d_e$.} 
\label{fig:den_temp}
\end{center}
\end{figure}

We focus our observations on the magnetic field at the front surface of the target, choosing the length of the target long enough that the back side does not influence the front (we have checked that the particles reflecting from the back do not reach the region $x_1 < 1150\,d_e$ where significant heating occurs until after $t = 2812.60\,\omega_p^{-1}$), and $n_e = 0 $ at the right wall to avoid significant particle loss at the boundary. We choose a step function at $x_1=1750  \,d_e$ to minimize the length and save computational time (see Fig.\,\ref{fig:den_temp}(a))\,.

Figure\,\ref{fig:den_temp} shows, in the simulation where $L_n = 400\,d_e$, that the laser produces temperature gradients that are not aligned with the density gradient associated with $L_n$. The laser enters the simulation domain from the left and at time $t \simeq 1200.50 \, \omega_p^{-1}$ penetrates the plasma up to $1000 \, d_e$ (Fig.\,\ref{fig:den_temp}(a)). The interaction of the laser with the plasma resonantly heats the electrons, consistent with the scaling of ref. \cite{Pukhov} (Fig.\,\ref{fig:den_temp}(b)). The temperature is defined as $T_e= \mathrm{Trace}(T_{ij})/3$, where $T_{ij} \equiv \int (u_i u_j/\gamma) f(u)d^3 u/ \int f(u)d^3 u$, calculated in the rest frame, is the temperature tensor, $u_i$ is the normalized proper velocity, $\gamma = \sqrt{1+ u^2}$, and $f(u)$ is the velocity distribution function.
By time $t \simeq 2812.60 \, \omega_p^{-1}$, the laser has created a conical shaped channel (see Fig.\,\ref{fig:den_temp}(c)) and induced a large thermal gradient with $L_T = 1000\,d_e$ pointing radially towards the axis of the laser beam (see Fig.\,\ref{fig:den_temp}(d)). The temperature gradient is not aligned with the density gradient along $x_1$ allowing the Biermann battery to generate a toroidal B-field. 

The average temperature along the line at $x_1 = 700 \, d_e$ is $\left<T_{e}\right>_{x2} = 0.34\,m_e c^2$ (see Fig.\,\ref{fig:den_temp}). Given this temperature and the maximum density $n_0=1.1\times 10^{22}\, cm^{-3}$, we conservatively estimate the collisionality. The ratio of $L_n$ to the electron collisional mean free path $l_e$ \cite{Huba2013}, $L_n/l_e = 0.00047 \ll 1$, therefore we neglect collisions.

\begin{figure}[htp]
\begin{center}
\includegraphics[scale=0.17]{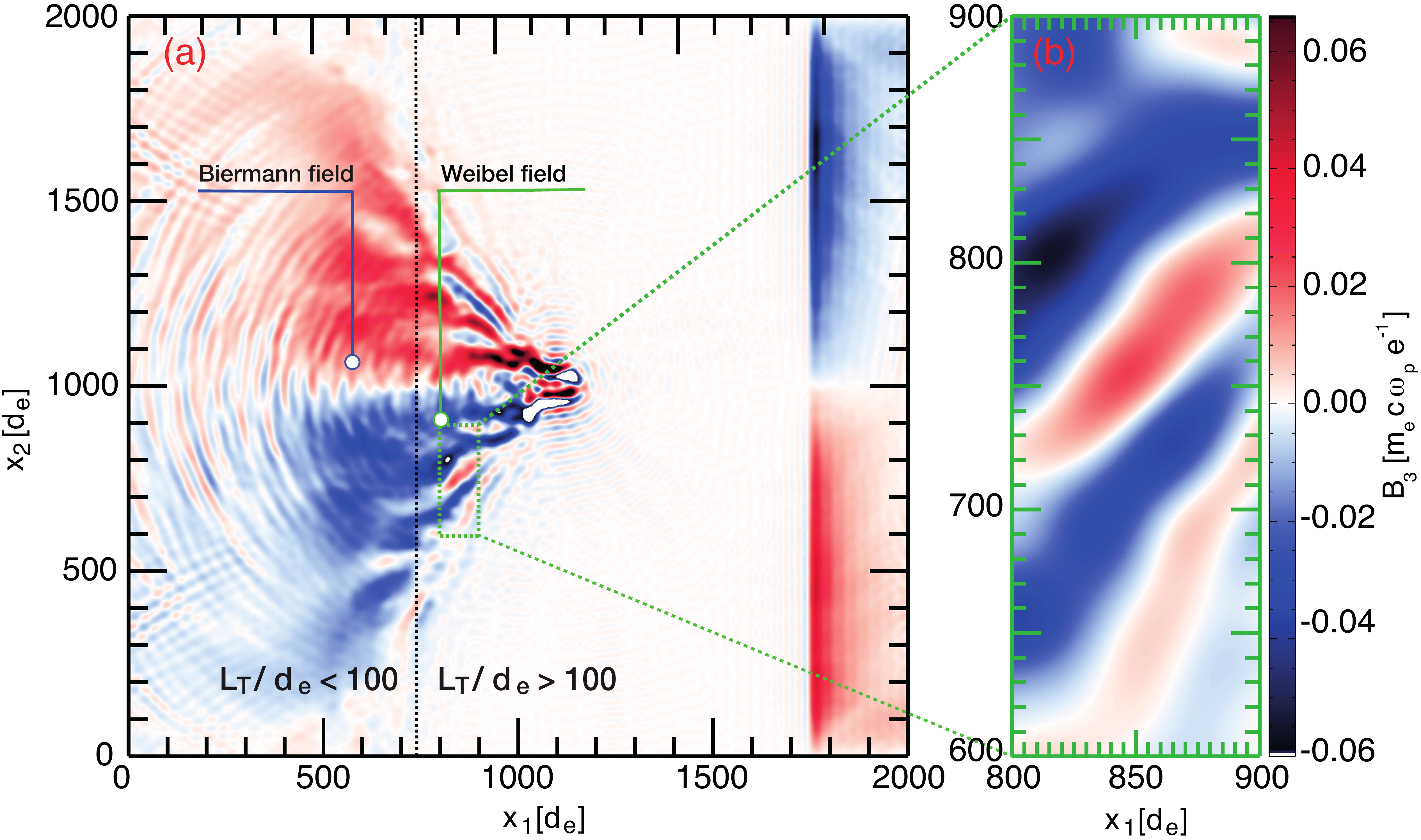}
\caption{(color online). (a) Out-of-plane magnetic field $B_3$ at $t = 2641.10 \,\omega_p^{-1} (440\,\text{fs})$ and (b) zoom-in of Weibel-generated magnetic filaments with $k \approx 0.06 \, d_e^{-1}$. The black dashed line in (a) indicates the transition point between the region where Biermann fields dominate ($L_T/d_e<100$) and the region where Weibel fields dominate ($L_T/d_e>100$).}
\label{fig:Mag_trans}
\end{center}
\end{figure}

Figure\,\ref{fig:Mag_trans} shows the Bierman-produced out-of-plane magnetic field $B_3$ at $t = 2641.10\,\omega_p^{-1}$ in the region $x_1 < \, 700 \, d_e$. However, alongside the Biermann-generated field, in the region $x_1 > \, 700 \, d_e$, a field due to the Weibel instability is also observed. The magnetic field reaches a maximum amplitude of the order of $0.065 \,m_e c/e \omega_p$ ($22 \,\text{MGauss}$). Note that a low-pass filter was applied to the magnetic field only allowing wavelengths above 31.4 $d_e$ (1.57 $\mu m$), mimicking the typical experimental resolution (see e.g. \cite{Wei2004}). The boundary between Biermann and Weibel regimes is estimated at the location where $L_T(x_1)/d_e(n_e(x_1)) \approx 100$ \cite{Schoeffler-2014,Schoeffler2016}, where $d_e(n_e(x_1))$ is the local electron inertial length. Remarkably, this transition occurs precisely at $x_1 = 700\, d_e$, indicated by the dotted vertical line in Fig.\,\ref{fig:Mag_trans}(a), as $d_e(n_e(x_1))=10\, d_e$ and $L_T(x_1)=1000\, d_e$ (see Fig.\,\ref{fig:den_temp}(c-d)).

\begin{figure}[htp]
\centering
\includegraphics[scale=0.18]{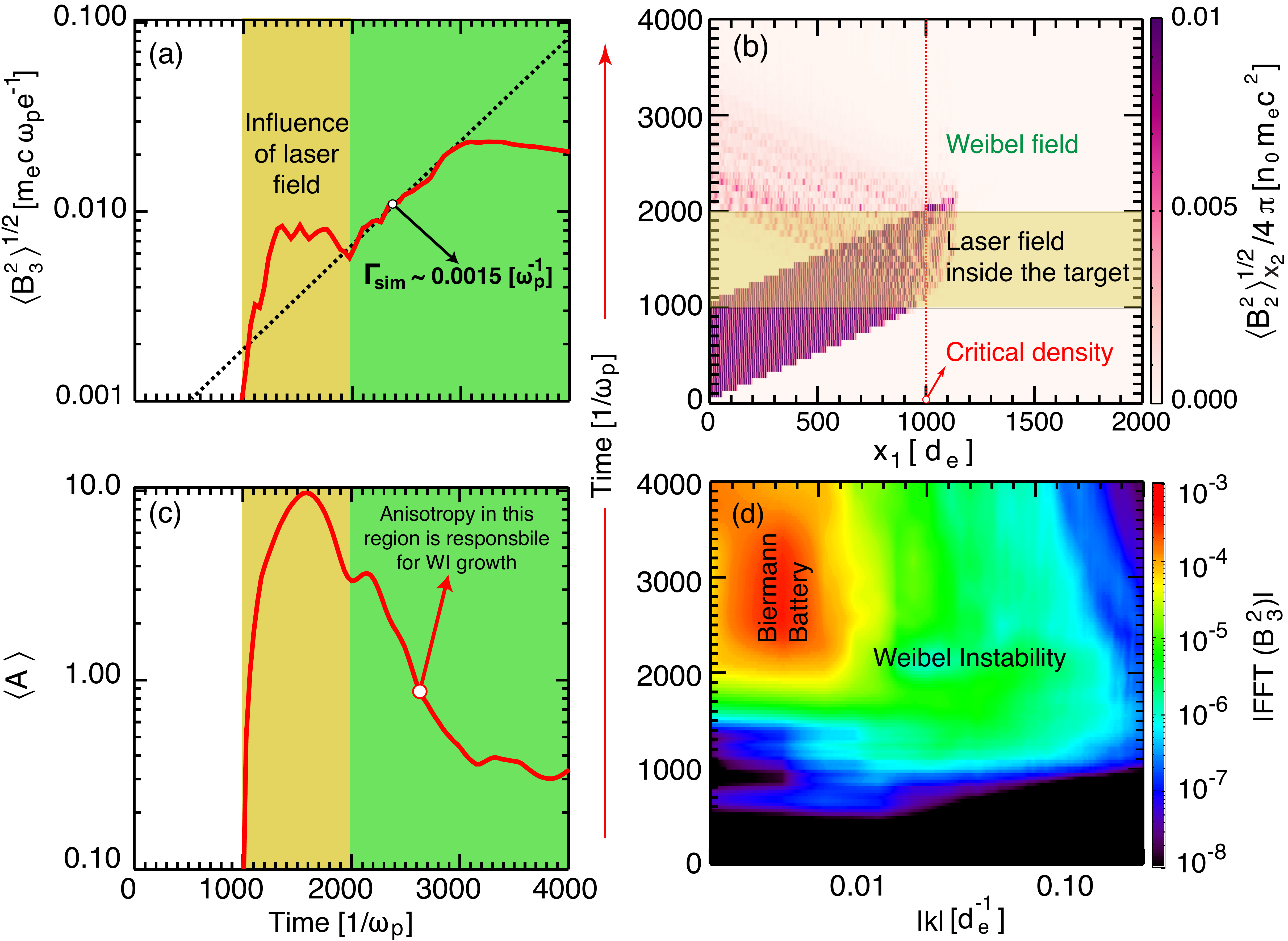}
\caption{(color online). (a) Temporal evolution of the square root of the average out-of-plane magnetic energy-density $\left<B_3^2\right>^{1/2}$ in the green box indicated in Fig\,\ref{fig:Mag_trans}. The slope of the curve in (a), identified as the Weibel growth rate, is $\simeq 0.0015 \, \omega_p$ (black dashed line). The average anisotropy $\left<A\right>$ in the green box indicated in Fig.\,\ref{fig:Mag_trans} is shown in (b). The temporal evolution of the magnetic field energy associated with the laser $\left<B_2^2\right>_{x_2}$, averaged along $x_2$,  as a function of $x_1$ is plotted in (c). The temporal evolution of the transverse magnetic field energy $B_3^2$ spectrum in (d) shows the contribution to B-field from the Weibel instability and the Biermann battery.}  
\label{fig:Growth_Ani_laser_fft}
\end{figure}

Figure\,\ref{fig:Growth_Ani_laser_fft}(a) shows the temporal evolution of the square root of the average out-of-plane magnetic energy-density $\left<B_3^2\right>^{1/2}$ in the region $x_1 = [800-900]\,d_e,x_2 =[600-900]\,d_e$, where the dominant source of the magnetic field is the Weibel instability. Between $2000-3000\, \omega_p^{-1}$, after the laser has passed this region (see Fig.\,\ref{fig:Growth_Ani_laser_fft}(b)), the laser magnetic fields are no longer present. Here, we observe an exponential growth of the magnetic field ($\Gamma_{sim} = 0.0015\, \omega_p$ with a corresponding wave-vector $k \simeq 0.15\,d_e^{-1}$, agreeing reasonably with theory from ref. \cite{Kaang}). The spatiotemporal evolution of the laser magnetic field energy shown in Fig.\,3 (b) shows that the end of the laser pulse passes the region where we calculate the growth rate ($x_1 < 900\,d_e$) at $t = 1950\,\omega_p^{-1}(\mathrm{322\,fs})$. Meanwhile, the expansion of the hot energetic electron population generated via laser-heating contributes to the average anisotropy in the velocity distribution (see Fig.\,\ref{fig:Growth_Ani_laser_fft}(c)) \cite{Schoeffler2017}. The anisotropy $A \equiv T_{hot}/T_{cold} - 1$, where $T_{hot}$ and $T_{cold}$ are the respective larger and smaller eigenvalues of the temperature tensor $T_{ij}$, provides the free energy that drives the Weibel instability. 

The time varying spectrum of $B_3^2$ in Fig.\,\ref{fig:Growth_Ani_laser_fft}(d) shows the contribution of the Weibel instability and the Biermann battery to the magnetic field energy. The spectra are obtained by performing a Fourier transform over the entire system for the out-of-plane magnetic fields, and then averaging over all directions of $\textbf{k}$. With the log scale it is not obvious that the energy contained in the Weibel magnetic fields is comparable to that of the Biermann. The Biermann magnetic field energy ($k d_e < 0.025 $) remains about five times higher than the Weibel magnetic fields energy ($k d_e > 0.025$) after $t = 2370\,\omega_p^{-1}$. 

\begin{figure}[h]
\begin{center}
	\includegraphics[scale=0.2]{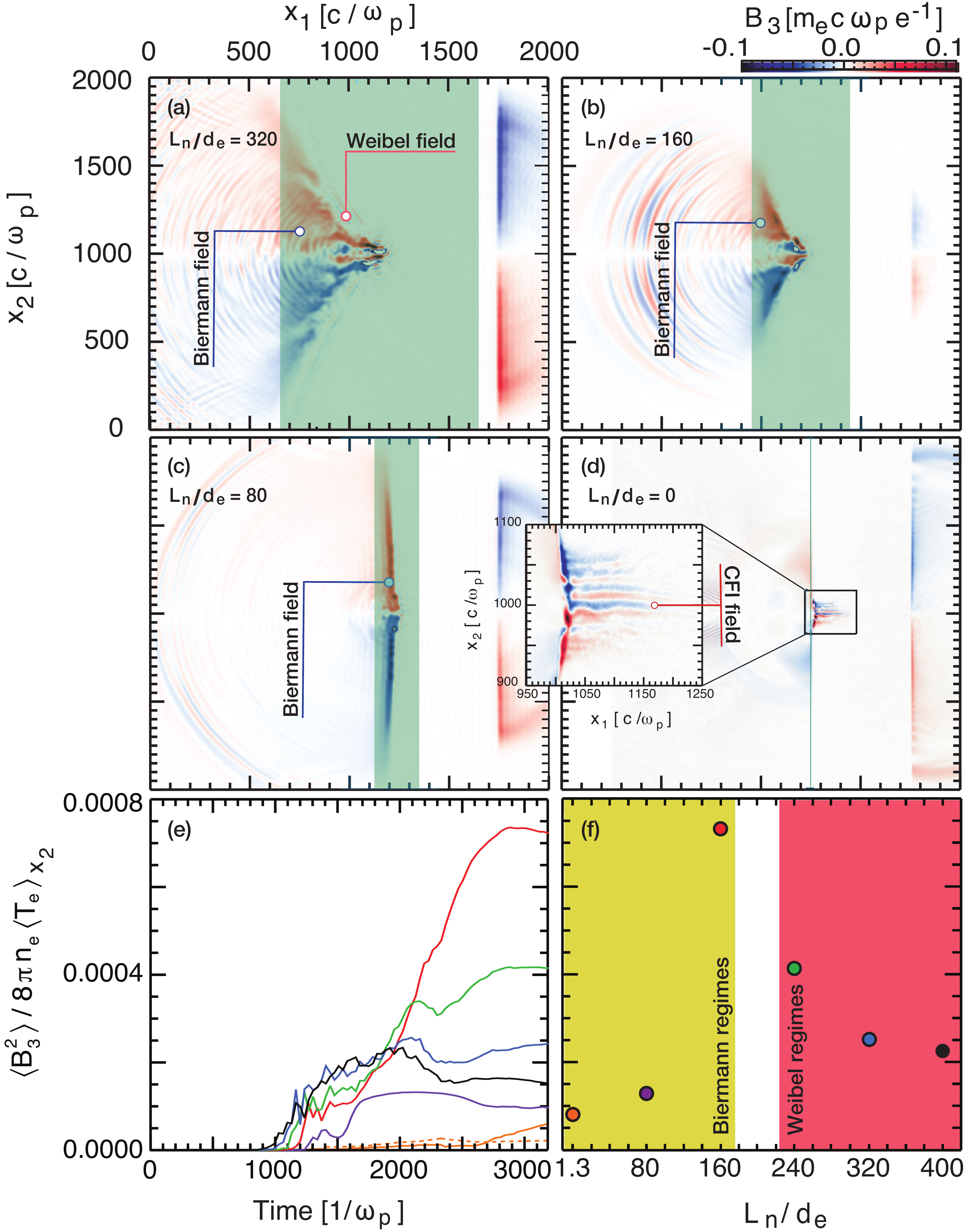}
	\caption{Out-of-plane magnetic field $B_3$ for $L_n/d_e= 320$(a), $160$(b), $80$(c) (with low-pass filter), and $0$(d) (without low-pass filter) at time $t = 2023.70\,\omega_p^{-1}$. Shaded regions  indicate where the mean field energy was averaged between $x_1$ $= 1250\,d_e - 1.875\, L_n$ and $1250\,d_e + 1.25\, L_n$. A zoom of the region where the current filamentation is found is included in (d). The temporal evolution of the average out-of-plane magnetic energy-density $\left<B_3^2\right>$ (averaged over the specified regions highlighted in (a)-(d) with low-pass filter) is shown in (e). The dashed line shows the average magnetic energy-density (without low-pass filter) in the range $x_1 = [950-1250]\,d_e$. The peak $\left<B_3^2\right>$ is plotted as a function of $L_n/d_e$ in (f). The Biermann field dominates over the Weibel where $L_n/d_e \geq 160$ (yellow region), while a region where the Weibel field dominates exists when $240 \geq L_n/d_e$ (red region).} 
	\label{fig:Field_com.eps}
\end{center}
\end{figure}

We performed a parameter scan for $\mathrm{L_n}/d_e = 0$, $80$, $160$, $240$, $320$, and $400$. Note that by the time the laser reaches the target at $t\sim 1250\,\omega_{p}^{-1}$, the length scale rises by $\sim \sqrt{k_B T_{e0}/m_i}\,t \sim 1.3\,d_e$, given $T_{e0}$ $= 1\,\mathrm{keV}$. Therefore, for $L_n/d_e = 0$, the effective density scale length is $1.3\, d_e$. Fig.\,\ref{fig:Field_com.eps}(a-d) shows $B_3$ at time $t = 2023.70\,\omega_p^{-1}$ (when the Weibel generated magnetic fields saturate in the $L_n/d_e = 400$ case, see Fig.\,\ref{fig:Field_com.eps}(e)) for a selection of $L_n/d_e$. With a target of sufficiently large $L_n/d_e > 160$, a region of Weibel generated magnetic fields is visible (see Fig.\,\ref{fig:Field_com.eps}(a) where $L_n/d_e = 320$). However, for $L_n/d_e \leq 160$, the Biermann magnetic field dominates, and no region exists where the Weibel instability is prominent (see Fig.\,\ref{fig:Field_com.eps}(b-d)).

Thin filaments in $B_3$ explained by the current filamentation instability (CFI) \cite{Fried, Kolodner1979, Shukla-JPP-2012} are observed in many experiments \cite{tata2002, Gode-2017} where a laser hits a plasma target with a sharp density profile. Fig.\,\ref{fig:Field_com.eps}(d) shows these filaments (without the low-pass filter). Unlike the Weibel generated field described in this work, a sharp relativistic electron beam provides the free energy rather than the thermal expansion of the plasma. In our simulations, the CFI field is much weaker than both the Weibel and Biermann fields for other $L_n/d_e$. Furthermore, in this Letter, we focus on the region with density and temperature gradients that lead to the Biermann battery and Weibel instability, rather than deep inside the target where these thin filamentary fields are found.

The magnetic energy-density produced from the laser-interaction depends on $L_n$. Fig.\,\ref{fig:Field_com.eps}(e) shows the temporal evolution of the average out-of-plane magnetic energy-density $\left<B_3^2\right>$ (with low-pass filter) in the region between $x_1 = 1250\,d_e - 1.875\,L_n$ and $x_1 = 1250\,d_e + 1.25\,L_n$ for each simulation (see highlighted regions in Fig.\,\ref{fig:Field_com.eps}(a-d)). Weibel fields are observed when $L_n/d_e > 160$, saturating at $t \sim 2000\, \omega_p^{-1}$. For all cases, the Biermann field grows and saturates after $t \geq 2150\, \omega_p^{-1}$. The dashed line shows $\left<B_3^2\right>$ (without low-pass filter) in the range $x_1 = [950-1250]\,d_e$ associated with the zoomed region in Fig.\,\ref{fig:Field_com.eps}(d), which peaks at $t \sim 2000\, \omega_p^{-1}$. This CFI magnetic field is much smaller than the dominant fields for bigger $L_n$. In Fig.\,\ref{fig:Field_com.eps}(f), the peak $\left<B_3^2\right>$ is shown as a function of $L_n/d_e$. The maximum $\left<B_3^2\right>$ occurs at $L_n/d_e = 160$, the transition between the Biermann and Weibel regimes.

The transition between the regimes where only the Biermann battery is present ($L_n/d_e < 160 $) and both the Weibel instability and the Biermann battery are present ($L_n/d_e > 160 $) can be probed experimentally. After the target is ionized by the pre-pulse, the plasma expands resulting in a non-uniform density with a gradient length that can reach several micrometers when the main pulse arrives. A possible model for the density scale length as a function of time yields: $L_n(t) = 14.5\,\mathrm{\mu m} \cdot \bar{I}_L^{10/27} \bar{A}^{-2/27} \bar{\Lambda}^{4/27} \bar{\lambda}_L^{4/9} \bar{\Delta t}^{31/27}$ \cite{PhysRevE.63.036410}, where the bar notation signifies quantities normalized to a typical pre-pulse laser with intensity of $I_L = 10^{12}\,\mathrm{W/cm^{-2}}$, the nuclear mass number $A\,= 2$, the Coulomb logarithm $\Lambda\,= 5$, the laser wavelength $\lambda_{0} = 1\,\mathrm{\mu m}$, and pulse delay $\Delta t = 200\,\mathrm{ps}$. For example, with these scalings, pulse delays $278\,\mathrm{ps}$ and $68.4\,\mathrm{ps}$ correspond with $L_n = 400\,d_e$ and $80\,d_e $, confirming the experimental feasibility of these density scale lengths. 

Magnetic fields can be measured using the synchrotron radiation in addition to the conventional method of proton radiography \cite{Li2007}. For the parameters of this study, radiation will have wavelength estimated between $190-1200\,\text{nm}$, while for higher power lasers, this signal would become stronger and approach x-ray frequencies. The detailed prediction of the radiation spectra, which can in principle be performed using radiation algorithms \cite{Martins, radio2018}, will be left for future work.

In this Letter, we have demonstrated the possibility to clearly observe the generation of electron Weibel magnetic fields in laboratory experiments. First-principles PIC simulations of the interaction of an intense laser pulse with an overdense plasma target have demostrated the Weibel instability in the presence of sufficiently weak gradients at the front of the target ($L_n/d_e \ge 160$ and $\text{w}_{FWHM} =100\,d_e$). The Weibel instability is driven by an electron pressure anisotropy caused by the rapid expansion of the electrons in the front of the target, following the laser-plasma interaction. The Weibel instability produces fields saturating at magnitudes comparable to the Biermann fields.

Finally, we note that density gradients needed to observe the instability at work could easily be achieved tuning the delay between the ionizing pre-pulse and the main pulse at existing laser facilities. For instance, facilities such as the Vulcan laser facility at Rutherford Appleton Laboratory \cite{Vulcan2017} with a peak intensity around $I_L = 10^{19}\,\mathrm{W/cm^2}$, wavelength $\lambda_{0} = 1.0 54\,\mu \text{m}$, a duration of hundreds of femtoseconds, and a contrast of $10^7$ would easily allow testing the interplay and the competition between the Weibel and the Biermann mechanisms.

\begin{acknowledgments}
This work was partially supported by the European Research Council (ERC-2015-InPairs-695088). Simulations were performed at the IST cluster (Lisbon, Portugal) and on the Marconi supercomputer (CINECA) in the framework of the HPC-Europa3 program.
\end{acknowledgments}

\bibliography{LaserTarget}

\begin{thebibliography}{48}%
\makeatletter
\providecommand \@ifxundefined [1]{%
 \@ifx{#1\undefined}
}%
\providecommand \@ifnum [1]{%
 \ifnum #1\expandafter \@firstoftwo
 \else \expandafter \@secondoftwo
 \fi
}%
\providecommand \@ifx [1]{%
 \ifx #1\expandafter \@firstoftwo
 \else \expandafter \@secondoftwo
 \fi
}%
\providecommand \natexlab [1]{#1}%
\providecommand \enquote  [1]{``#1''}%
\providecommand \bibnamefont  [1]{#1}%
\providecommand \bibfnamefont [1]{#1}%
\providecommand \citenamefont [1]{#1}%
\providecommand \href@noop [0]{\@secondoftwo}%
\providecommand \href [0]{\begingroup \@sanitize@url \@href}%
\providecommand \@href[1]{\@@startlink{#1}\@@href}%
\providecommand \@@href[1]{\endgroup#1\@@endlink}%
\providecommand \@sanitize@url [0]{\catcode `\\12\catcode `\$12\catcode
  `\&12\catcode `\#12\catcode `\^12\catcode `\_12\catcode `\%12\relax}%
\providecommand \@@startlink[1]{}%
\providecommand \@@endlink[0]{}%
\providecommand \url  [0]{\begingroup\@sanitize@url \@url }%
\providecommand \@url [1]{\endgroup\@href {#1}{\urlprefix }}%
\providecommand \urlprefix  [0]{URL }%
\providecommand \Eprint [0]{\href }%
\providecommand \doibase [0]{http://dx.doi.org/}%
\providecommand \selectlanguage [0]{\@gobble}%
\providecommand \bibinfo  [0]{\@secondoftwo}%
\providecommand \bibfield  [0]{\@secondoftwo}%
\providecommand \translation [1]{[#1]}%
\providecommand \BibitemOpen [0]{}%
\providecommand \bibitemStop [0]{}%
\providecommand \bibitemNoStop [0]{.\EOS\space}%
\providecommand \EOS [0]{\spacefactor3000\relax}%
\providecommand \BibitemShut  [1]{\csname bibitem#1\endcsname}%
\let\auto@bib@innerbib\@empty
\bibitem [{\citenamefont {Uzdensky}\ and\ \citenamefont
  {Rightley}(2014)}]{Uzdensky-2014}%
  \BibitemOpen
  \bibfield  {author} {\bibinfo {author} {\bibfnamefont {D.~A.}\ \bibnamefont
  {Uzdensky}}\ and\ \bibinfo {author} {\bibfnamefont {S.}~\bibnamefont
  {Rightley}},\ }\href@noop {} {\bibfield  {journal} {\bibinfo  {journal} {Rep.
  Prog. Phys.}\ }\textbf {\bibinfo {volume} {77}},\ \bibinfo {pages} {036902}
  (\bibinfo {year} {2014})}\BibitemShut {NoStop}%
\bibitem [{\citenamefont {{L. W. Widrow}}(2002)}]{Widrow}%
  \BibitemOpen
  \bibfield  {author} {\bibinfo {author} {\bibnamefont {{L. W. Widrow}}},\
  }\href@noop {} {\bibfield  {journal} {\bibinfo  {journal} {Rev. Mod. Phys.}\
  }\textbf {\bibinfo {volume} {74}},\ \bibinfo {pages} {775} (\bibinfo {year}
  {2002})}\BibitemShut {NoStop}%
\bibitem [{\citenamefont {{P. P. Kronberg}}(2002)}]{Kronberg}%
  \BibitemOpen
  \bibfield  {author} {\bibinfo {author} {\bibnamefont {{P. P. Kronberg}}},\
  }\href@noop {} {\bibfield  {journal} {\bibinfo  {journal} {Phys. Today}\
  }\textbf {\bibinfo {volume} {55}},\ \bibinfo {pages} {1240} (\bibinfo {year}
  {2002})}\BibitemShut {NoStop}%
\bibitem [{\citenamefont {Kulsrud}\ and\ \citenamefont
  {Zweibel}(2008)}]{Kulsrud-2008}%
  \BibitemOpen
  \bibfield  {author} {\bibinfo {author} {\bibfnamefont {R.~M.}\ \bibnamefont
  {Kulsrud}}\ and\ \bibinfo {author} {\bibfnamefont {E.~G.}\ \bibnamefont
  {Zweibel}},\ }\href@noop {} {\bibfield  {journal} {\bibinfo  {journal} {Rep.
  Prog. Phys.}\ }\textbf {\bibinfo {volume} {71}},\ \bibinfo {pages} {046901}
  (\bibinfo {year} {2008})}\BibitemShut {NoStop}%
\bibitem [{\citenamefont {Silva}\ \emph {et~al.}(2002)\citenamefont {Silva},
  \citenamefont {Fonseca}, \citenamefont {Tonge}, \citenamefont {Mori},\ and\
  \citenamefont {Dawson}}]{Luis-2002}%
  \BibitemOpen
  \bibfield  {author} {\bibinfo {author} {\bibfnamefont {L.~O.}\ \bibnamefont
  {Silva}}, \bibinfo {author} {\bibfnamefont {R.~A.}\ \bibnamefont {Fonseca}},
  \bibinfo {author} {\bibfnamefont {J.~W.}\ \bibnamefont {Tonge}}, \bibinfo
  {author} {\bibfnamefont {W.~B.}\ \bibnamefont {Mori}}, \ and\ \bibinfo
  {author} {\bibfnamefont {J.~M.}\ \bibnamefont {Dawson}},\ }\href@noop {}
  {\bibfield  {journal} {\bibinfo  {journal} {Phys. Plasmas}\ }\textbf
  {\bibinfo {volume} {9}},\ \bibinfo {pages} {2458} (\bibinfo {year}
  {2002})}\BibitemShut {NoStop}%
\bibitem [{\citenamefont {Tzoufras}\ \emph {et~al.}(2006)\citenamefont
  {Tzoufras}, \citenamefont {Ren}, \citenamefont {Tsung}, \citenamefont
  {Tonge}, \citenamefont {Mori}, \citenamefont {Fiore}, \citenamefont
  {Fonseca},\ and\ \citenamefont {Silva}}]{Tzoufras-2006}%
  \BibitemOpen
  \bibfield  {author} {\bibinfo {author} {\bibfnamefont {M.}~\bibnamefont
  {Tzoufras}}, \bibinfo {author} {\bibfnamefont {C.}~\bibnamefont {Ren}},
  \bibinfo {author} {\bibfnamefont {F.~S.}\ \bibnamefont {Tsung}}, \bibinfo
  {author} {\bibfnamefont {J.~W.}\ \bibnamefont {Tonge}}, \bibinfo {author}
  {\bibfnamefont {W.~B.}\ \bibnamefont {Mori}}, \bibinfo {author}
  {\bibfnamefont {M.}~\bibnamefont {Fiore}}, \bibinfo {author} {\bibfnamefont
  {R.~A.}\ \bibnamefont {Fonseca}}, \ and\ \bibinfo {author} {\bibfnamefont
  {L.~O.}\ \bibnamefont {Silva}},\ }\href@noop {} {\bibfield  {journal}
  {\bibinfo  {journal} {Phys. Rev. Lett.}\ }\textbf {\bibinfo {volume} {96}},\
  \bibinfo {pages} {105002} (\bibinfo {year} {2006})}\BibitemShut {NoStop}%
\bibitem [{\citenamefont {Shukla}\ \emph {et~al.}(2012)\citenamefont {Shukla},
  \citenamefont {Stockem}, \citenamefont {Fiuza},\ and\ \citenamefont
  {Silva}}]{Shukla-JPP-2012}%
  \BibitemOpen
  \bibfield  {author} {\bibinfo {author} {\bibfnamefont {N.}~\bibnamefont
  {Shukla}}, \bibinfo {author} {\bibfnamefont {A.}~\bibnamefont {Stockem}},
  \bibinfo {author} {\bibfnamefont {F.}~\bibnamefont {Fiuza}}, \ and\ \bibinfo
  {author} {\bibfnamefont {L.~O.}\ \bibnamefont {Silva}},\ }\href@noop {}
  {\bibfield  {journal} {\bibinfo  {journal} {J. Plasma Phys.}\ }\textbf
  {\bibinfo {volume} {78}},\ \bibinfo {pages} {181} (\bibinfo {year}
  {2012})}\BibitemShut {NoStop}%
\bibitem [{\citenamefont {{M. V. Medvedev and A.
  Leob}}(1999)}]{Medvedev-ApJ-1999}%
  \BibitemOpen
  \bibfield  {author} {\bibinfo {author} {\bibnamefont {{M. V. Medvedev and A.
  Leob}}},\ }\href@noop {} {\bibfield  {journal} {\bibinfo  {journal}
  {Astrophys. J.}\ }\textbf {\bibinfo {volume} {526}},\ \bibinfo {pages} {697}
  (\bibinfo {year} {1999})}\BibitemShut {NoStop}%
\bibitem [{\citenamefont {Fonseca}\ \emph
  {et~al.}(2002{\natexlab{a}})\citenamefont {Fonseca}, \citenamefont {Silva},
  \citenamefont {Tonge}, \citenamefont {Hemker}, \citenamefont {Dawson},\ and\
  \citenamefont {Mori}}]{Ricardo-POP-1999}%
  \BibitemOpen
  \bibfield  {author} {\bibinfo {author} {\bibfnamefont {R.~A.}\ \bibnamefont
  {Fonseca}}, \bibinfo {author} {\bibfnamefont {L.~O.}\ \bibnamefont {Silva}},
  \bibinfo {author} {\bibfnamefont {J.}~\bibnamefont {Tonge}}, \bibinfo
  {author} {\bibfnamefont {R.~G.}\ \bibnamefont {Hemker}}, \bibinfo {author}
  {\bibfnamefont {J.~M.}\ \bibnamefont {Dawson}}, \ and\ \bibinfo {author}
  {\bibfnamefont {W.~B.}\ \bibnamefont {Mori}},\ }\href@noop {} {\bibfield
  {journal} {\bibinfo  {journal} {IEEE Trans. Plasma Sci.}\ }\textbf {\bibinfo
  {volume} {30}},\ \bibinfo {pages} {28} (\bibinfo {year}
  {2002}{\natexlab{a}})}\BibitemShut {NoStop}%
\bibitem [{\citenamefont {Medvedev}\ \emph {et~al.}(2004)\citenamefont
  {Medvedev}, \citenamefont {Fiore}, \citenamefont {Fonseca}, \citenamefont
  {Silva},\ and\ \citenamefont {Mori}}]{Medvedev-ApJ-2005}%
  \BibitemOpen
  \bibfield  {author} {\bibinfo {author} {\bibfnamefont {M.~V.}\ \bibnamefont
  {Medvedev}}, \bibinfo {author} {\bibfnamefont {M.}~\bibnamefont {Fiore}},
  \bibinfo {author} {\bibfnamefont {R.~A.}\ \bibnamefont {Fonseca}}, \bibinfo
  {author} {\bibfnamefont {L.~O.}\ \bibnamefont {Silva}}, \ and\ \bibinfo
  {author} {\bibfnamefont {W.~B.}\ \bibnamefont {Mori}},\ }\href@noop {}
  {\bibfield  {journal} {\bibinfo  {journal} {Astrophys. J.}\ }\textbf
  {\bibinfo {volume} {618}},\ \bibinfo {pages} {L75} (\bibinfo {year}
  {2004})}\BibitemShut {NoStop}%
\bibitem [{\citenamefont {Nishikawa}\ \emph {et~al.}(2009)\citenamefont
  {Nishikawa}, \citenamefont {Niemiec}, \citenamefont {Hardee}, \citenamefont
  {Medvedev}, \citenamefont {Sol}, \citenamefont {Mizuno}, \citenamefont
  {Zhang}, \citenamefont {Pohl}, \citenamefont {Oka},\ and\ \citenamefont
  {Hartmann}}]{Nishikawa-2009}%
  \BibitemOpen
  \bibfield  {author} {\bibinfo {author} {\bibfnamefont {K.~I.}\ \bibnamefont
  {Nishikawa}}, \bibinfo {author} {\bibfnamefont {J.}~\bibnamefont {Niemiec}},
  \bibinfo {author} {\bibfnamefont {P.~E.}\ \bibnamefont {Hardee}}, \bibinfo
  {author} {\bibfnamefont {M.}~\bibnamefont {Medvedev}}, \bibinfo {author}
  {\bibfnamefont {H.}~\bibnamefont {Sol}}, \bibinfo {author} {\bibfnamefont
  {Y.}~\bibnamefont {Mizuno}}, \bibinfo {author} {\bibfnamefont
  {B.}~\bibnamefont {Zhang}}, \bibinfo {author} {\bibfnamefont
  {M.}~\bibnamefont {Pohl}}, \bibinfo {author} {\bibfnamefont {M.}~\bibnamefont
  {Oka}}, \ and\ \bibinfo {author} {\bibfnamefont {D.~H.}\ \bibnamefont
  {Hartmann}},\ }\href@noop {} {\bibfield  {journal} {\bibinfo  {journal}
  {{Astrophys. J. Lett.}}\ }\textbf {\bibinfo {volume} {698}},\ \bibinfo
  {pages} {L10} (\bibinfo {year} {2009})}\BibitemShut {NoStop}%
\bibitem [{\citenamefont {{N. Shukla and P.K. Shukla}}(2010)}]{Shukla2010}%
  \BibitemOpen
  \bibfield  {author} {\bibinfo {author} {\bibnamefont {{N. Shukla and P.K.
  Shukla}}},\ }\href@noop {} {\bibfield  {journal} {\bibinfo  {journal} {J.
  Plasma Phys.}\ }\textbf {\bibinfo {volume} {76}},\ \bibinfo {pages} {1}
  (\bibinfo {year} {2010})}\BibitemShut {NoStop}%
\bibitem [{\citenamefont {Bret}(2009)}]{Bret-APJ-2009}%
  \BibitemOpen
  \bibfield  {author} {\bibinfo {author} {\bibfnamefont {A.}~\bibnamefont
  {Bret}},\ }\href@noop {} {\bibfield  {journal} {\bibinfo  {journal}
  {Astrophys. J.}\ }\textbf {\bibinfo {volume} {699}},\ \bibinfo {pages} {990}
  (\bibinfo {year} {2009})}\BibitemShut {NoStop}%
\bibitem [{\citenamefont {Ruyer}\ \emph {et~al.}(2016)\citenamefont {Ruyer},
  \citenamefont {Gremillet}, \citenamefont {Bonnaud},\ and\ \citenamefont
  {Riconda}}]{Ruyer-PRL-2017}%
  \BibitemOpen
  \bibfield  {author} {\bibinfo {author} {\bibfnamefont {C.}~\bibnamefont
  {Ruyer}}, \bibinfo {author} {\bibfnamefont {L.}~\bibnamefont {Gremillet}},
  \bibinfo {author} {\bibfnamefont {G.}~\bibnamefont {Bonnaud}}, \ and\
  \bibinfo {author} {\bibfnamefont {C.}~\bibnamefont {Riconda}},\ }\href@noop
  {} {\bibfield  {journal} {\bibinfo  {journal} {Physical Review Letters}\
  }\textbf {\bibinfo {volume} {117}},\ \bibinfo {pages} {065001} (\bibinfo
  {year} {2016})}\BibitemShut {NoStop}%
\bibitem [{\citenamefont {Gruzinov}(2001)}]{Gruzinov-2001}%
  \BibitemOpen
  \bibfield  {author} {\bibinfo {author} {\bibfnamefont {A.}~\bibnamefont
  {Gruzinov}},\ }\href@noop {} {\bibfield  {journal} {\bibinfo  {journal}
  {{Astrophys. J. Lett.}}\ }\textbf {\bibinfo {volume} {563}},\ \bibinfo
  {pages} {L15} (\bibinfo {year} {2001})}\BibitemShut {NoStop}%
\bibitem [{\citenamefont {Brandenburg}\ \emph {et~al.}(2012)\citenamefont
  {Brandenburg}, \citenamefont {Sokoloff},\ and\ \citenamefont
  {Subramanian}}]{Brandenburg2012}%
  \BibitemOpen
  \bibfield  {author} {\bibinfo {author} {\bibfnamefont {A.}~\bibnamefont
  {Brandenburg}}, \bibinfo {author} {\bibfnamefont {D.}~\bibnamefont
  {Sokoloff}}, \ and\ \bibinfo {author} {\bibfnamefont {K.}~\bibnamefont
  {Subramanian}},\ }\href@noop {} {\bibfield  {journal} {\bibinfo  {journal}
  {{Space Sci. Rev.}}\ }\textbf {\bibinfo {volume} {169}},\ \bibinfo {pages}
  {123} (\bibinfo {year} {2012})}\BibitemShut {NoStop}%
\bibitem [{\citenamefont {Weibel}(1959)}]{Weibel-1959}%
  \BibitemOpen
  \bibfield  {author} {\bibinfo {author} {\bibfnamefont {E.~S.}\ \bibnamefont
  {Weibel}},\ }\href@noop {} {\bibfield  {journal} {\bibinfo  {journal} {Phys.
  Rev. Lett.}\ }\textbf {\bibinfo {volume} {2}},\ \bibinfo {pages} {83}
  (\bibinfo {year} {1959})}\BibitemShut {NoStop}%
\bibitem [{\citenamefont {Bret}\ \emph {et~al.}(2010)\citenamefont {Bret},
  \citenamefont {Gremillet},\ and\ \citenamefont {Dieckmann}}]{Bret-POP-2010}%
  \BibitemOpen
  \bibfield  {author} {\bibinfo {author} {\bibfnamefont {A.}~\bibnamefont
  {Bret}}, \bibinfo {author} {\bibfnamefont {L.}~\bibnamefont {Gremillet}}, \
  and\ \bibinfo {author} {\bibfnamefont {M.~E.}\ \bibnamefont {Dieckmann}},\
  }\href@noop {} {\bibfield  {journal} {\bibinfo  {journal} {Physics of
  Plasmas}\ }\textbf {\bibinfo {volume} {17}},\ \bibinfo {pages} {120501}
  (\bibinfo {year} {2010})}\BibitemShut {NoStop}%
\bibitem [{\citenamefont {Tatarakis}\ \emph
  {et~al.}(2002{\natexlab{a}})\citenamefont {Tatarakis}, \citenamefont {Watts},
  \citenamefont {Beg}, \citenamefont {Clark}, \citenamefont {Dangor},
  \citenamefont {Gopal}, \citenamefont {Haines}, \citenamefont {Norreys},
  \citenamefont {Wagner}, \citenamefont {Wei}, \citenamefont {Zepf},\ and\
  \citenamefont {Krushelnick}}]{Tata-2002}%
  \BibitemOpen
  \bibfield  {author} {\bibinfo {author} {\bibfnamefont {M.}~\bibnamefont
  {Tatarakis}}, \bibinfo {author} {\bibfnamefont {I.}~\bibnamefont {Watts}},
  \bibinfo {author} {\bibfnamefont {F.~N.}\ \bibnamefont {Beg}}, \bibinfo
  {author} {\bibfnamefont {E.~L.}\ \bibnamefont {Clark}}, \bibinfo {author}
  {\bibfnamefont {A.~E.}\ \bibnamefont {Dangor}}, \bibinfo {author}
  {\bibfnamefont {A.}~\bibnamefont {Gopal}}, \bibinfo {author} {\bibfnamefont
  {M.~G.}\ \bibnamefont {Haines}}, \bibinfo {author} {\bibfnamefont {P.~A.}\
  \bibnamefont {Norreys}}, \bibinfo {author} {\bibfnamefont {U.}~\bibnamefont
  {Wagner}}, \bibinfo {author} {\bibfnamefont {M.-S.}\ \bibnamefont {Wei}},
  \bibinfo {author} {\bibfnamefont {M.}~\bibnamefont {Zepf}}, \ and\ \bibinfo
  {author} {\bibfnamefont {K.}~\bibnamefont {Krushelnick}},\ }\href@noop {}
  {\bibfield  {journal} {\bibinfo  {journal} {Nature}\ }\textbf {\bibinfo
  {volume} {415}},\ \bibinfo {pages} {280} (\bibinfo {year}
  {2002}{\natexlab{a}})}\BibitemShut {NoStop}%
\bibitem [{\citenamefont {Mondal}\ \emph {et~al.}(2012)\citenamefont {Mondal},
  \citenamefont {Narayanan}, \citenamefont {Ding}, \citenamefont {Lad},
  \citenamefont {Hao}, \citenamefont {Ahmad}, \citenamefont {Wang},
  \citenamefont {Sheng}, \citenamefont {Sengupta}, \citenamefont {Kaw},
  \citenamefont {Das},\ and\ \citenamefont {Kumar}}]{Mondal}%
  \BibitemOpen
  \bibfield  {author} {\bibinfo {author} {\bibfnamefont {S.}~\bibnamefont
  {Mondal}}, \bibinfo {author} {\bibfnamefont {V.}~\bibnamefont {Narayanan}},
  \bibinfo {author} {\bibfnamefont {W.~J.}\ \bibnamefont {Ding}}, \bibinfo
  {author} {\bibfnamefont {A.~D.}\ \bibnamefont {Lad}}, \bibinfo {author}
  {\bibfnamefont {B.}~\bibnamefont {Hao}}, \bibinfo {author} {\bibfnamefont
  {S.}~\bibnamefont {Ahmad}}, \bibinfo {author} {\bibfnamefont {W.~M.}\
  \bibnamefont {Wang}}, \bibinfo {author} {\bibfnamefont {Z.~M.}\ \bibnamefont
  {Sheng}}, \bibinfo {author} {\bibfnamefont {S.}~\bibnamefont {Sengupta}},
  \bibinfo {author} {\bibfnamefont {P.}~\bibnamefont {Kaw}}, \bibinfo {author}
  {\bibfnamefont {A.}~\bibnamefont {Das}}, \ and\ \bibinfo {author}
  {\bibfnamefont {G.~R.}\ \bibnamefont {Kumar}},\ }\href@noop {} {\bibfield
  {journal} {\bibinfo  {journal} {P. Natl. Acad. Sci. USA}\ }\textbf {\bibinfo
  {volume} {109}},\ \bibinfo {pages} {8011} (\bibinfo {year}
  {2012})}\BibitemShut {NoStop}%
\bibitem [{\citenamefont {Ruyer}\ \emph {et~al.}(2015)\citenamefont {Ruyer},
  \citenamefont {Gremillet},\ and\ \citenamefont {Bonnaud}}]{Ruyer-2015}%
  \BibitemOpen
  \bibfield  {author} {\bibinfo {author} {\bibfnamefont {C.}~\bibnamefont
  {Ruyer}}, \bibinfo {author} {\bibfnamefont {L.}~\bibnamefont {Gremillet}}, \
  and\ \bibinfo {author} {\bibfnamefont {G.}~\bibnamefont {Bonnaud}},\
  }\href@noop {} {\bibfield  {journal} {\bibinfo  {journal} {Phys. Plasmas}\
  }\textbf {\bibinfo {volume} {22}},\ \bibinfo {pages} {082107} (\bibinfo
  {year} {2015})}\BibitemShut {NoStop}%
\bibitem [{\citenamefont {Fiuza}\ \emph {et~al.}(2012)\citenamefont {Fiuza},
  \citenamefont {Fonseca}, \citenamefont {Tonge}, \citenamefont {Mori},\ and\
  \citenamefont {Silva}}]{Fiuza-2012}%
  \BibitemOpen
  \bibfield  {author} {\bibinfo {author} {\bibfnamefont {F.}~\bibnamefont
  {Fiuza}}, \bibinfo {author} {\bibfnamefont {R.~A.}\ \bibnamefont {Fonseca}},
  \bibinfo {author} {\bibfnamefont {J.}~\bibnamefont {Tonge}}, \bibinfo
  {author} {\bibfnamefont {W.~B.}\ \bibnamefont {Mori}}, \ and\ \bibinfo
  {author} {\bibfnamefont {L.~O.}\ \bibnamefont {Silva}},\ }\href@noop {}
  {\bibfield  {journal} {\bibinfo  {journal} {Phys. Rev. Lett.}\ }\textbf
  {\bibinfo {volume} {108}},\ \bibinfo {pages} {235004} (\bibinfo {year}
  {2012})}\BibitemShut {NoStop}%
\bibitem [{\citenamefont {Boella}\ \emph {et~al.}(2018)\citenamefont {Boella},
  \citenamefont {Fi{\'{u}}za}, \citenamefont {Novo}, \citenamefont {Fonseca},\
  and\ \citenamefont {Silva}}]{Boella_2018}%
  \BibitemOpen
  \bibfield  {author} {\bibinfo {author} {\bibfnamefont {E.}~\bibnamefont
  {Boella}}, \bibinfo {author} {\bibfnamefont {F.}~\bibnamefont {Fi{\'{u}}za}},
  \bibinfo {author} {\bibfnamefont {A.~S.}\ \bibnamefont {Novo}}, \bibinfo
  {author} {\bibfnamefont {R.~A.}\ \bibnamefont {Fonseca}}, \ and\ \bibinfo
  {author} {\bibfnamefont {L.~O.}\ \bibnamefont {Silva}},\ }\href@noop {}
  {\bibfield  {journal} {\bibinfo  {journal} {Plasma Phys. C. F.}\ }\textbf
  {\bibinfo {volume} {60}},\ \bibinfo {pages} {035010} (\bibinfo {year}
  {2018})}\BibitemShut {NoStop}%
\bibitem [{\citenamefont {Shukla}\ \emph {et~al.}(2009)\citenamefont {Shukla},
  \citenamefont {Shukla},\ and\ \citenamefont {Stenflo}}]{PhysRevE.80.027401}%
  \BibitemOpen
  \bibfield  {author} {\bibinfo {author} {\bibfnamefont {N.}~\bibnamefont
  {Shukla}}, \bibinfo {author} {\bibfnamefont {P.~K.}\ \bibnamefont {Shukla}},
  \ and\ \bibinfo {author} {\bibfnamefont {L.}~\bibnamefont {Stenflo}},\
  }\href@noop {} {\bibfield  {journal} {\bibinfo  {journal} {Phys. Rev. E}\
  }\textbf {\bibinfo {volume} {80}},\ \bibinfo {pages} {027401} (\bibinfo
  {year} {2009})}\BibitemShut {NoStop}%
\bibitem [{\citenamefont {Stamper}\ \emph {et~al.}(1971)\citenamefont
  {Stamper}, \citenamefont {Papadopoulos}, \citenamefont {Sudan}, \citenamefont
  {Dean}, \citenamefont {McLean},\ and\ \citenamefont {Dawson}}]{Stamper-1971}%
  \BibitemOpen
  \bibfield  {author} {\bibinfo {author} {\bibfnamefont {J.~A.}\ \bibnamefont
  {Stamper}}, \bibinfo {author} {\bibfnamefont {K.}~\bibnamefont
  {Papadopoulos}}, \bibinfo {author} {\bibfnamefont {R.~N.}\ \bibnamefont
  {Sudan}}, \bibinfo {author} {\bibfnamefont {S.~O.}\ \bibnamefont {Dean}},
  \bibinfo {author} {\bibfnamefont {E.~A.}\ \bibnamefont {McLean}}, \ and\
  \bibinfo {author} {\bibfnamefont {J.~M.}\ \bibnamefont {Dawson}},\
  }\href@noop {} {\bibfield  {journal} {\bibinfo  {journal} {Phys. Rev. Lett.}\
  }\textbf {\bibinfo {volume} {26}},\ \bibinfo {pages} {1012} (\bibinfo {year}
  {1971})}\BibitemShut {NoStop}%
\bibitem [{\citenamefont {Sakagami}\ \emph {et~al.}(1979)\citenamefont
  {Sakagami}, \citenamefont {Kawakami}, \citenamefont {Nagao},\ and\
  \citenamefont {Yamanaka}}]{Sakagami1979}%
  \BibitemOpen
  \bibfield  {author} {\bibinfo {author} {\bibfnamefont {Y.}~\bibnamefont
  {Sakagami}}, \bibinfo {author} {\bibfnamefont {H.}~\bibnamefont {Kawakami}},
  \bibinfo {author} {\bibfnamefont {S.}~\bibnamefont {Nagao}}, \ and\ \bibinfo
  {author} {\bibfnamefont {C.}~\bibnamefont {Yamanaka}},\ }\href@noop {}
  {\bibfield  {journal} {\bibinfo  {journal} {Phys. Rev. Lett.}\ }\textbf
  {\bibinfo {volume} {42}},\ \bibinfo {pages} {839} (\bibinfo {year}
  {1979})}\BibitemShut {NoStop}%
\bibitem [{\citenamefont {Cadjan}\ \emph {et~al.}(1997)\citenamefont {Cadjan},
  \citenamefont {Ivanov},\ and\ \citenamefont
  {Ivlev}}]{cadjan_ivanov_ivlev_1997}%
  \BibitemOpen
  \bibfield  {author} {\bibinfo {author} {\bibfnamefont {M.~G.}\ \bibnamefont
  {Cadjan}}, \bibinfo {author} {\bibfnamefont {M.~F.}\ \bibnamefont {Ivanov}},
  \ and\ \bibinfo {author} {\bibfnamefont {A.~V.}\ \bibnamefont {Ivlev}},\
  }\href@noop {} {\bibfield  {journal} {\bibinfo  {journal} {Laser Par. Beams}\
  }\textbf {\bibinfo {volume} {15}},\ \bibinfo {pages} {33} (\bibinfo {year}
  {1997})}\BibitemShut {NoStop}%
\bibitem [{\citenamefont {Wilks}\ and\ \citenamefont
  {Kruer}(1997)}]{Wilks1997}%
  \BibitemOpen
  \bibfield  {author} {\bibinfo {author} {\bibfnamefont {S.~C.}\ \bibnamefont
  {Wilks}}\ and\ \bibinfo {author} {\bibfnamefont {W.~L.}\ \bibnamefont
  {Kruer}},\ }\href@noop {} {\bibfield  {journal} {\bibinfo  {journal} {IEEE J.
  Quantum Elect.}\ }\textbf {\bibinfo {volume} {33}},\ \bibinfo {pages} {1954}
  (\bibinfo {year} {1997})}\BibitemShut {NoStop}%
\bibitem [{\citenamefont {Schoeffler}\ \emph {et~al.}(2014)\citenamefont
  {Schoeffler}, \citenamefont {Loureiro}, \citenamefont {Fonseca},\ and\
  \citenamefont {Silva}}]{Schoeffler-2014}%
  \BibitemOpen
  \bibfield  {author} {\bibinfo {author} {\bibfnamefont {K.~M.}\ \bibnamefont
  {Schoeffler}}, \bibinfo {author} {\bibfnamefont {N.~F.}\ \bibnamefont
  {Loureiro}}, \bibinfo {author} {\bibfnamefont {R.~A.}\ \bibnamefont
  {Fonseca}}, \ and\ \bibinfo {author} {\bibfnamefont {L.~O.}\ \bibnamefont
  {Silva}},\ }\href@noop {} {\bibfield  {journal} {\bibinfo  {journal} {Phys.
  Rev. Lett.}\ }\textbf {\bibinfo {volume} {112}},\ \bibinfo {pages} {175001}
  (\bibinfo {year} {2014})}\BibitemShut {NoStop}%
\bibitem [{\citenamefont {Schoeffler}\ \emph {et~al.}(2016)\citenamefont
  {Schoeffler}, \citenamefont {Loureiro}, \citenamefont {Fonseca},\ and\
  \citenamefont {Silva}}]{Schoeffler2016}%
  \BibitemOpen
  \bibfield  {author} {\bibinfo {author} {\bibfnamefont {K.~M.}\ \bibnamefont
  {Schoeffler}}, \bibinfo {author} {\bibfnamefont {N.~F.}\ \bibnamefont
  {Loureiro}}, \bibinfo {author} {\bibfnamefont {R.~A.}\ \bibnamefont
  {Fonseca}}, \ and\ \bibinfo {author} {\bibfnamefont {L.~O.}\ \bibnamefont
  {Silva}},\ }\href@noop {} {\bibfield  {journal} {\bibinfo  {journal} {Phys.
  Plasmas}\ }\textbf {\bibinfo {volume} {23}},\ \bibinfo {pages} {056304}
  (\bibinfo {year} {2016})}\BibitemShut {NoStop}%
\bibitem [{\citenamefont {Fonseca}\ \emph
  {et~al.}(2002{\natexlab{b}})\citenamefont {Fonseca}, \citenamefont {Silva},
  \citenamefont {Tsung}, \citenamefont {Decyk}, \citenamefont {Lu},
  \citenamefont {Ren}, \citenamefont {Mori}, \citenamefont {Deng},
  \citenamefont {Lee}, \citenamefont {Katsouleas},\ and\ \citenamefont
  {Adam}}]{Fonseca-2002}%
  \BibitemOpen
  \bibfield  {author} {\bibinfo {author} {\bibfnamefont {R.~A.}\ \bibnamefont
  {Fonseca}}, \bibinfo {author} {\bibfnamefont {L.~O.}\ \bibnamefont {Silva}},
  \bibinfo {author} {\bibfnamefont {F.~S.}\ \bibnamefont {Tsung}}, \bibinfo
  {author} {\bibfnamefont {V.~K.}\ \bibnamefont {Decyk}}, \bibinfo {author}
  {\bibfnamefont {W.}~\bibnamefont {Lu}}, \bibinfo {author} {\bibfnamefont
  {C.}~\bibnamefont {Ren}}, \bibinfo {author} {\bibfnamefont {W.~B.}\
  \bibnamefont {Mori}}, \bibinfo {author} {\bibfnamefont {S.}~\bibnamefont
  {Deng}}, \bibinfo {author} {\bibfnamefont {S.}~\bibnamefont {Lee}}, \bibinfo
  {author} {\bibfnamefont {T.}~\bibnamefont {Katsouleas}}, \ and\ \bibinfo
  {author} {\bibfnamefont {J.~C.}\ \bibnamefont {Adam}},\ }\href@noop {}
  {\bibfield  {journal} {\bibinfo  {journal} {Lect. Notes Comput. Sci.}\
  }\textbf {\bibinfo {volume} {2331}},\ \bibinfo {pages} {046401} (\bibinfo
  {year} {2002}{\natexlab{b}})}\BibitemShut {NoStop}%
\bibitem [{\citenamefont {Fonseca}\ \emph {et~al.}(2008)\citenamefont
  {Fonseca}, \citenamefont {Martins}, \citenamefont {Silva}, \citenamefont
  {Tonge}, \citenamefont {Tsung},\ and\ \citenamefont {Mori}}]{Fonseca2008}%
  \BibitemOpen
  \bibfield  {author} {\bibinfo {author} {\bibfnamefont {R.~A.}\ \bibnamefont
  {Fonseca}}, \bibinfo {author} {\bibfnamefont {S.~F.}\ \bibnamefont
  {Martins}}, \bibinfo {author} {\bibfnamefont {L.~O.}\ \bibnamefont {Silva}},
  \bibinfo {author} {\bibfnamefont {J.~W.}\ \bibnamefont {Tonge}}, \bibinfo
  {author} {\bibfnamefont {F.~S.}\ \bibnamefont {Tsung}}, \ and\ \bibinfo
  {author} {\bibfnamefont {W.~B.}\ \bibnamefont {Mori}},\ }\href@noop {}
  {\bibfield  {journal} {\bibinfo  {journal} {Plasma Phys. C. F.}\ }\textbf
  {\bibinfo {volume} {50}},\ \bibinfo {pages} {124034} (\bibinfo {year}
  {2008})}\BibitemShut {NoStop}%
\bibitem [{\citenamefont {Fonseca}\ \emph {et~al.}(2013)\citenamefont
  {Fonseca}, \citenamefont {Vieira}, \citenamefont {Fiuza}, \citenamefont
  {Davidson}, \citenamefont {Tsung}, \citenamefont {Mori},\ and\ \citenamefont
  {Silva}}]{Fonseca2013}%
  \BibitemOpen
  \bibfield  {author} {\bibinfo {author} {\bibfnamefont {R.~A.}\ \bibnamefont
  {Fonseca}}, \bibinfo {author} {\bibfnamefont {J.}~\bibnamefont {Vieira}},
  \bibinfo {author} {\bibfnamefont {F.}~\bibnamefont {Fiuza}}, \bibinfo
  {author} {\bibfnamefont {A.}~\bibnamefont {Davidson}}, \bibinfo {author}
  {\bibfnamefont {F.~S.}\ \bibnamefont {Tsung}}, \bibinfo {author}
  {\bibfnamefont {W.~B.}\ \bibnamefont {Mori}}, \ and\ \bibinfo {author}
  {\bibfnamefont {L.~O.}\ \bibnamefont {Silva}},\ }\href@noop {} {\bibfield
  {journal} {\bibinfo  {journal} {Plasma Phys. C. F.}\ }\textbf {\bibinfo
  {volume} {55}},\ \bibinfo {pages} {124011} (\bibinfo {year}
  {2013})}\BibitemShut {NoStop}%
\bibitem [{\citenamefont {Chopineau}\ \emph {et~al.}(2019)\citenamefont
  {Chopineau}, \citenamefont {Leblanc}, \citenamefont {Blaclard}, \citenamefont
  {Denoeud}, \citenamefont {Th\'evenet}, \citenamefont {Vay}, \citenamefont
  {Bonnaud}, \citenamefont {Martin}, \citenamefont {Vincenti},\ and\
  \citenamefont {Qu\'er\'e}}]{Chopineau}%
  \BibitemOpen
  \bibfield  {author} {\bibinfo {author} {\bibfnamefont {L.}~\bibnamefont
  {Chopineau}}, \bibinfo {author} {\bibfnamefont {A.}~\bibnamefont {Leblanc}},
  \bibinfo {author} {\bibfnamefont {G.}~\bibnamefont {Blaclard}}, \bibinfo
  {author} {\bibfnamefont {A.}~\bibnamefont {Denoeud}}, \bibinfo {author}
  {\bibfnamefont {M.}~\bibnamefont {Th\'evenet}}, \bibinfo {author}
  {\bibfnamefont {J.-L.}\ \bibnamefont {Vay}}, \bibinfo {author} {\bibfnamefont
  {G.}~\bibnamefont {Bonnaud}}, \bibinfo {author} {\bibfnamefont
  {P.}~\bibnamefont {Martin}}, \bibinfo {author} {\bibfnamefont
  {H.}~\bibnamefont {Vincenti}}, \ and\ \bibinfo {author} {\bibfnamefont
  {F.}~\bibnamefont {Qu\'er\'e}},\ }\href@noop {} {\bibfield  {journal}
  {\bibinfo  {journal} {Phys. Rev. X}\ }\textbf {\bibinfo {volume} {9}},\
  \bibinfo {pages} {011050} (\bibinfo {year} {2019})}\BibitemShut {NoStop}%
\bibitem [{\citenamefont {Wei}\ \emph {et~al.}(2004)\citenamefont {Wei},
  \citenamefont {Beg}, \citenamefont {Clark}, \citenamefont {Dangor},
  \citenamefont {Evans}, \citenamefont {Gopal}, \citenamefont {Ledingham},
  \citenamefont {McKenna}, \citenamefont {Norreys}, \citenamefont {Tatarakis},
  \citenamefont {Zepf},\ and\ \citenamefont {Krushelnick}}]{Wei2004}%
  \BibitemOpen
  \bibfield  {author} {\bibinfo {author} {\bibfnamefont {M.~S.}\ \bibnamefont
  {Wei}}, \bibinfo {author} {\bibfnamefont {F.~N.}\ \bibnamefont {Beg}},
  \bibinfo {author} {\bibfnamefont {E.~L.}\ \bibnamefont {Clark}}, \bibinfo
  {author} {\bibfnamefont {A.~E.}\ \bibnamefont {Dangor}}, \bibinfo {author}
  {\bibfnamefont {R.~G.}\ \bibnamefont {Evans}}, \bibinfo {author}
  {\bibfnamefont {A.}~\bibnamefont {Gopal}}, \bibinfo {author} {\bibfnamefont
  {K.~W.~D.}\ \bibnamefont {Ledingham}}, \bibinfo {author} {\bibfnamefont
  {P.}~\bibnamefont {McKenna}}, \bibinfo {author} {\bibfnamefont {P.~A.}\
  \bibnamefont {Norreys}}, \bibinfo {author} {\bibfnamefont {M.}~\bibnamefont
  {Tatarakis}}, \bibinfo {author} {\bibfnamefont {M.}~\bibnamefont {Zepf}}, \
  and\ \bibinfo {author} {\bibfnamefont {K.}~\bibnamefont {Krushelnick}},\
  }\href@noop {} {\bibfield  {journal} {\bibinfo  {journal} {Phys. Rev. E}\
  }\textbf {\bibinfo {volume} {70}},\ \bibinfo {pages} {056412} (\bibinfo
  {year} {2004})}\BibitemShut {NoStop}%
\bibitem [{\citenamefont {Pukhov}\ \emph {et~al.}(1999)\citenamefont {Pukhov},
  \citenamefont {Sheng},\ and\ \citenamefont {ter Vehn}}]{Pukhov}%
  \BibitemOpen
  \bibfield  {author} {\bibinfo {author} {\bibfnamefont {A.}~\bibnamefont
  {Pukhov}}, \bibinfo {author} {\bibfnamefont {Z.-M.}\ \bibnamefont {Sheng}}, \
  and\ \bibinfo {author} {\bibfnamefont {J.~M.}\ \bibnamefont {ter Vehn}},\
  }\href@noop {} {\bibfield  {journal} {\bibinfo  {journal} {Phys. Plasmas}\
  }\textbf {\bibinfo {volume} {6}},\ \bibinfo {pages} {2847} (\bibinfo {year}
  {1999})}\BibitemShut {NoStop}%
\bibitem [{\citenamefont {Huba}(2013)}]{Huba2013}%
  \BibitemOpen
  \bibfield  {author} {\bibinfo {author} {\bibfnamefont {J.~D.}\ \bibnamefont
  {Huba}},\ }\href@noop {} {\emph {\bibinfo {title} {Plasma Physics}}}\
  (\bibinfo  {publisher} {Naval Research Laboratory},\ \bibinfo {address}
  {Washington, DC},\ \bibinfo {year} {2013})\BibitemShut {NoStop}%
\bibitem [{\citenamefont {Kaang}\ \emph {et~al.}(2009)\citenamefont {Kaang},
  \citenamefont {Ryu},\ and\ \citenamefont {Yoon}}]{Kaang}%
  \BibitemOpen
  \bibfield  {author} {\bibinfo {author} {\bibfnamefont {H.~H.}\ \bibnamefont
  {Kaang}}, \bibinfo {author} {\bibfnamefont {C.-M.}\ \bibnamefont {Ryu}}, \
  and\ \bibinfo {author} {\bibfnamefont {P.~H.}\ \bibnamefont {Yoon}},\
  }\href@noop {} {\bibfield  {journal} {\bibinfo  {journal} {Phys. Plasmas}\
  }\textbf {\bibinfo {volume} {16}},\ \bibinfo {pages} {082103} (\bibinfo
  {year} {2009})}\BibitemShut {NoStop}%
\bibitem [{\citenamefont {Schoeffler}\ \emph {et~al.}(2018)\citenamefont
  {Schoeffler}, \citenamefont {Loureiro},\ and\ \citenamefont
  {Silva}}]{Schoeffler2017}%
  \BibitemOpen
  \bibfield  {author} {\bibinfo {author} {\bibfnamefont {K.~M.}\ \bibnamefont
  {Schoeffler}}, \bibinfo {author} {\bibfnamefont {N.~F.}\ \bibnamefont
  {Loureiro}}, \ and\ \bibinfo {author} {\bibfnamefont {L.~O.}\ \bibnamefont
  {Silva}},\ }\href@noop {} {\bibfield  {journal} {\bibinfo  {journal} {Phys.
  Rev. E}\ }\textbf {\bibinfo {volume} {97}},\ \bibinfo {pages} {033204}
  (\bibinfo {year} {2018})}\BibitemShut {NoStop}%
\bibitem [{\citenamefont {Fried}(1959)}]{Fried}%
  \BibitemOpen
  \bibfield  {author} {\bibinfo {author} {\bibfnamefont {B.~D.}\ \bibnamefont
  {Fried}},\ }\href@noop {} {\bibfield  {journal} {\bibinfo  {journal} {Phys.
  Fluids}\ }\textbf {\bibinfo {volume} {2}},\ \bibinfo {pages} {337} (\bibinfo
  {year} {1959})}\BibitemShut {NoStop}%
\bibitem [{\citenamefont {Kolodner}\ and\ \citenamefont
  {Yablonovitch}(1979)}]{Kolodner1979}%
  \BibitemOpen
  \bibfield  {author} {\bibinfo {author} {\bibfnamefont {P.}~\bibnamefont
  {Kolodner}}\ and\ \bibinfo {author} {\bibfnamefont {E.}~\bibnamefont
  {Yablonovitch}},\ }\href@noop {} {\bibfield  {journal} {\bibinfo  {journal}
  {Phys. Rev. Lett.}\ }\textbf {\bibinfo {volume} {43}},\ \bibinfo {pages}
  {1402} (\bibinfo {year} {1979})}\BibitemShut {NoStop}%
\bibitem [{\citenamefont {Tatarakis}\ \emph
  {et~al.}(2002{\natexlab{b}})\citenamefont {Tatarakis}, \citenamefont {Gopal},
  \citenamefont {Watts}, \citenamefont {Beg}, \citenamefont {Dangor},
  \citenamefont {Krushelnick}, \citenamefont {Wagner}, \citenamefont {Norreys},
  \citenamefont {Clark}, \citenamefont {Zepf},\ and\ \citenamefont
  {Evans}}]{tata2002}%
  \BibitemOpen
  \bibfield  {author} {\bibinfo {author} {\bibfnamefont {M.}~\bibnamefont
  {Tatarakis}}, \bibinfo {author} {\bibfnamefont {A.}~\bibnamefont {Gopal}},
  \bibinfo {author} {\bibfnamefont {I.}~\bibnamefont {Watts}}, \bibinfo
  {author} {\bibfnamefont {F.~N.}\ \bibnamefont {Beg}}, \bibinfo {author}
  {\bibfnamefont {A.~E.}\ \bibnamefont {Dangor}}, \bibinfo {author}
  {\bibfnamefont {K.}~\bibnamefont {Krushelnick}}, \bibinfo {author}
  {\bibfnamefont {U.}~\bibnamefont {Wagner}}, \bibinfo {author} {\bibfnamefont
  {P.~A.}\ \bibnamefont {Norreys}}, \bibinfo {author} {\bibfnamefont {E.~L.}\
  \bibnamefont {Clark}}, \bibinfo {author} {\bibfnamefont {M.}~\bibnamefont
  {Zepf}}, \ and\ \bibinfo {author} {\bibfnamefont {R.~G.}\ \bibnamefont
  {Evans}},\ }\href@noop {} {\bibfield  {journal} {\bibinfo  {journal} {Phys.
  Plasmas}\ }\textbf {\bibinfo {volume} {9}},\ \bibinfo {pages} {2244}
  (\bibinfo {year} {2002}{\natexlab{b}})}\BibitemShut {NoStop}%
\bibitem [{\citenamefont {G\"ode}\ \emph {et~al.}(2017)\citenamefont {G\"ode},
  \citenamefont {R\"odel}, \citenamefont {Zeil}, \citenamefont {Mishra},
  \citenamefont {Gauthier}, \citenamefont {Brack}, \citenamefont {Kluge},
  \citenamefont {MacDonald}, \citenamefont {Metzkes}, \citenamefont {Obst},
  \citenamefont {Rehwald}, \citenamefont {Ruyer}, \citenamefont {Schlenvoigt},
  \citenamefont {Schumaker}, \citenamefont {Sommer}, \citenamefont {Cowan},
  \citenamefont {Schramm}, \citenamefont {Glenzer},\ and\ \citenamefont
  {Fiuza}}]{Gode-2017}%
  \BibitemOpen
  \bibfield  {author} {\bibinfo {author} {\bibfnamefont {S.}~\bibnamefont
  {G\"ode}}, \bibinfo {author} {\bibfnamefont {C.}~\bibnamefont {R\"odel}},
  \bibinfo {author} {\bibfnamefont {K.}~\bibnamefont {Zeil}}, \bibinfo {author}
  {\bibfnamefont {R.}~\bibnamefont {Mishra}}, \bibinfo {author} {\bibfnamefont
  {M.}~\bibnamefont {Gauthier}}, \bibinfo {author} {\bibfnamefont {F.-E.}\
  \bibnamefont {Brack}}, \bibinfo {author} {\bibfnamefont {T.}~\bibnamefont
  {Kluge}}, \bibinfo {author} {\bibfnamefont {M.~J.}\ \bibnamefont
  {MacDonald}}, \bibinfo {author} {\bibfnamefont {J.}~\bibnamefont {Metzkes}},
  \bibinfo {author} {\bibfnamefont {L.}~\bibnamefont {Obst}}, \bibinfo {author}
  {\bibfnamefont {M.}~\bibnamefont {Rehwald}}, \bibinfo {author} {\bibfnamefont
  {C.}~\bibnamefont {Ruyer}}, \bibinfo {author} {\bibfnamefont {H.-P.}\
  \bibnamefont {Schlenvoigt}}, \bibinfo {author} {\bibfnamefont
  {W.}~\bibnamefont {Schumaker}}, \bibinfo {author} {\bibfnamefont
  {P.}~\bibnamefont {Sommer}}, \bibinfo {author} {\bibfnamefont {T.~E.}\
  \bibnamefont {Cowan}}, \bibinfo {author} {\bibfnamefont {U.}~\bibnamefont
  {Schramm}}, \bibinfo {author} {\bibfnamefont {S.}~\bibnamefont {Glenzer}}, \
  and\ \bibinfo {author} {\bibfnamefont {F.}~\bibnamefont {Fiuza}},\
  }\href@noop {} {\bibfield  {journal} {\bibinfo  {journal} {Phys. Rev. Lett.}\
  }\textbf {\bibinfo {volume} {118}},\ \bibinfo {pages} {194801} (\bibinfo
  {year} {2017})}\BibitemShut {NoStop}%
\bibitem [{\citenamefont {Li}\ and\ \citenamefont
  {Zhang}(2001)}]{PhysRevE.63.036410}%
  \BibitemOpen
  \bibfield  {author} {\bibinfo {author} {\bibfnamefont {Y.~J.}\ \bibnamefont
  {Li}}\ and\ \bibinfo {author} {\bibfnamefont {J.}~\bibnamefont {Zhang}},\
  }\href@noop {} {\bibfield  {journal} {\bibinfo  {journal} {Phys. Rev. E}\
  }\textbf {\bibinfo {volume} {63}},\ \bibinfo {pages} {036410} (\bibinfo
  {year} {2001})}\BibitemShut {NoStop}%
\bibitem [{\citenamefont {Li}\ \emph {et~al.}(2007)\citenamefont {Li},
  \citenamefont {S\'eguin}, \citenamefont {Frenje}, \citenamefont {Rygg},
  \citenamefont {Petrasso}, \citenamefont {Town}, \citenamefont {Landen},
  \citenamefont {Knauer},\ and\ \citenamefont {Smalyuk}}]{Li2007}%
  \BibitemOpen
  \bibfield  {author} {\bibinfo {author} {\bibfnamefont {C.~K.}\ \bibnamefont
  {Li}}, \bibinfo {author} {\bibfnamefont {F.~H.}\ \bibnamefont {S\'eguin}},
  \bibinfo {author} {\bibfnamefont {J.~A.}\ \bibnamefont {Frenje}}, \bibinfo
  {author} {\bibfnamefont {J.~R.}\ \bibnamefont {Rygg}}, \bibinfo {author}
  {\bibfnamefont {R.~D.}\ \bibnamefont {Petrasso}}, \bibinfo {author}
  {\bibfnamefont {R.~P.~J.}\ \bibnamefont {Town}}, \bibinfo {author}
  {\bibfnamefont {O.~L.}\ \bibnamefont {Landen}}, \bibinfo {author}
  {\bibfnamefont {J.~P.}\ \bibnamefont {Knauer}}, \ and\ \bibinfo {author}
  {\bibfnamefont {V.~A.}\ \bibnamefont {Smalyuk}},\ }\href@noop {} {\bibfield
  {journal} {\bibinfo  {journal} {Phys. Rev. Lett.}\ }\textbf {\bibinfo
  {volume} {99}},\ \bibinfo {pages} {055001} (\bibinfo {year}
  {2007})}\BibitemShut {NoStop}%
\bibitem [{\citenamefont {Martins}\ \emph {et~al.}(2009)\citenamefont
  {Martins}, \citenamefont {Martins}, \citenamefont {Fonseca},\ and\
  \citenamefont {Silva}}]{Martins}%
  \BibitemOpen
  \bibfield  {author} {\bibinfo {author} {\bibfnamefont {J.~L.}\ \bibnamefont
  {Martins}}, \bibinfo {author} {\bibfnamefont {S.~F.}\ \bibnamefont
  {Martins}}, \bibinfo {author} {\bibfnamefont {R.~A.}\ \bibnamefont
  {Fonseca}}, \ and\ \bibinfo {author} {\bibfnamefont {L.~O.}\ \bibnamefont
  {Silva}},\ }\href@noop {} {\bibfield  {journal} {\bibinfo  {journal} {Proc.
  SPIE}\ }\textbf {\bibinfo {volume} {7359}},\ \bibinfo {pages} {73590V}
  (\bibinfo {year} {2009})}\BibitemShut {NoStop}%
\bibitem [{\citenamefont {{M. Pardal, A. Sainte-Marie, A. Reboul-Salze, J.
  Vieira, and R. A. Fonseca}}(2018)}]{radio2018}%
  \BibitemOpen
  \bibfield  {author} {\bibinfo {author} {\bibnamefont {{M. Pardal, A.
  Sainte-Marie, A. Reboul-Salze, J. Vieira, and R. A. Fonseca}}},\ }in\
  \href@noop {} {\emph {\bibinfo {booktitle} {45th EPS Conference on Plasma
  Physics}}}\ (\bibinfo {year} {2018})\BibitemShut {NoStop}%
\bibitem [{Vul()}]{Vulcan2017}%
  \BibitemOpen
  \href@noop {} {}\bibinfo {howpublished}
  {\url{www.clf.stfc.ac.uk/Pages/Vulcan-laser.aspx}}\BibitemShut {NoStop}%
\end{thebibliography}%


%

\end{document}